\documentclass[aps,prd
,preprint,tightenlines,superscriptaddress,nofootinbib,showpacs]{revtex4}
\usepackage{amssymb,latexsym}
\usepackage{amsmath,amsbsy,bbm}
\usepackage{epsfig,bm}
\usepackage{graphicx,comment}
\DeclareMathOperator{\tr}{tr}

\begin{document}
\def\a{{\alpha}}
\def\b{{\beta}}
\def\d{{\delta}}
\def\D{{\Delta}}
\def\e{{\varepsilon}}
\def\g{{\gamma}}
\def\G{{\Gamma}}
\def\k{{\kappa}}
\def\l{{\lambda}}
\def\L{{\Lambda}}
\def\m{{\mu}}
\def\n{{\nu}}
\def\o{{\omega}}
\def\O{{\Omega}}
\def\S{{\Sigma}}
\def\s{{\sigma}}
\def\th{{\theta}}

\def\ol#1{{\overline{#1}}}

\def\Dslash{D\hskip-0.65em /}
\def\Dtslash{\tilde{D} \hskip-0.65em /}

\def\CPT{{$\chi$PT}}
\def\QCPT{{Q$\chi$PT}}
\def\PQCPT{{PQ$\chi$PT}}
\def\tr{\text{tr}}
\def\str{\text{str}}
\def\diag{\text{diag}}
\def\order{{\mathcal O}}

\def\cC{{\mathcal C}}
\def\cB{{\mathcal B}}
\def\cT{{\mathcal T}}
\def\cQ{{\mathcal Q}}
\def\cL{{\mathcal L}}
\def\cO{{\mathcal O}}
\def\cA{{\mathcal A}}
\def\cQ{{\mathcal Q}}
\def\cR{{\mathcal R}}
\def\cH{{\mathcal H}}
\def\cW{{\mathcal W}}
\def\cM{{\mathcal M}}
\def\cD{{\mathcal D}}
\def\cN{{\mathcal N}}
\def\cP{{\mathcal P}}
\def\cK{{\mathcal K}}
\def\Qt{{\tilde{Q}}}
\def\Dt{{\tilde{D}}}
\def\St{{\tilde{\Sigma}}}
\def\cBt{{\tilde{\mathcal{B}}}}
\def\cDt{{\tilde{\mathcal{D}}}}
\def\cTt{{\tilde{\mathcal{T}}}}
\def\cMt{{\tilde{\mathcal{M}}}}
\def\At{{\tilde{A}}}
\def\cNt{{\tilde{\mathcal{N}}}}
\def\cOt{{\tilde{\mathcal{O}}}}
\def\cPt{{\tilde{\mathcal{P}}}}
\def\cI{{\mathcal{I}}}
\def\cJ{{\mathcal{J}}}

\def\eqref#1{{(\ref{#1})}}

\title{Flavor Twisted Boundary Conditions, 
Pion Momentum, and the Pion Electromagnetic Form Factor}

\author{F.-J.~Jiang}
\email[]{fjjiang@itp.unibe.ch}
\affiliation{Department of Physics, Duke University, Box 90305, Durham, NC 27708-0305, USA}
\affiliation{Institute for Theoretical Physics, Bern University, Sidlerstrasse 5, CH-3012 Bern, Switzerland}
\author{B.~C.~Tiburzi}
\email[]{bctiburz@phy.duke.edu}
\affiliation{Department of Physics, Duke University, Box 90305, Durham, NC 27708-0305, USA}

\date{\today}


\begin{abstract}
We investigate the utility of partially twisted boundary 
conditions in lattice calculations of meson observables. 
For dynamical simulations, we show that the pion
dispersion relation is modified by volume effects. 
In the isospin limit, we demonstrate that the pion electromagnetic 
form factor can be computed on the lattice at continuous values of the 
momentum transfer.
Furthermore, the finite volume effects are under theoretical control
for extraction of the pion charge radius. 
\end{abstract}
\maketitle

\section{Introduction}

Lattice QCD simulations enable first principles calculation of hadronic properties. Currently, 
however, observables calculated on the lattice cannot be directly compared with nature
due to systematic error. 
This error arises from approximations used in solving the 
theory numerically, such as the finite extent of the lattice, 
the non-vanishing lattice spacing, 
and larger than physical quark masses.  
For a large class of observables, effective field theories (EFTs) can be tailored to 
describe the approximations used in lattice QCD.
These EFTs are the reliable tool to remove systematic error 
in extrapolating data to the physical point, see, e.g.~\cite{Bernard:2002yk}. 
With growing computational resources and
improved numerical algorithms, we are beginning to enter a period
in which lattice data in conjunction with EFT
methods will enable first principles predictions.

A further limitation encountered in lattice simulations is the available momentum. 
With periodic boundary conditions, the momentum modes are quantized. On 
currently available dynamical lattices, the lowest non-vanishing momentum 
mode is $\sim 500 \, \texttt{MeV}$. While EFTs can be tailored to address
systematic error in lattice QCD, these low-energy theories cannot be utilized
to predict the momentum dependence outside the range of their applicability. 
The extraction of quantities requiring non-zero momentum, such as phase shifts, electromagnetic 
moments and radii, is thus limited by coarse grained lattice momentum.
Spatially periodic boundary conditions, however, are not mandated.
So-called twisted boundary conditions~%
\cite{Gross:1982at,Roberge:1986mm,Wiese:1991ku,Luscher:1996sc,Bucarelli:1998mu,Guagnelli:2003hw,Kiskis:2002gr,Kiskis:2003rd,Kim:2002np,Kim:2003xt} 
have gained recent attention because they can be utilized to produce continuous hadron momentum~%
\cite{Bedaque:2004kc,deDivitiis:2004kq,Sachrajda:2004mi,Bedaque:2004ax,Tiburzi:2005hg,Flynn:2005in,Guadagnoli:2005be,Aarts:2006wt,Tiburzi:2006px}.

The electromagnetic properties of hadrons paint an intuitive 
picture of the charge and magnetism distributions of quarks 
confined in strongly interacting bound states.
To resolve such properties of hadrons, one requires 
momentum transfer and this presents a challenge for the lattice.  
Calculations of the pion form factor from lattice QCD, for example, 
have matured since the original pioneering works~%
\cite{Martinelli:1988bh,Draper:1989bp},
but still suffer from systematic error. 
Recently it was demonstrated~%
\cite{Bunton:2006va} 
that current dynamical calculations~%
\cite{Bonnet:2004fr,Hashimoto:2005am,Brommel:2006ww} 
of the pion charge radii are limited by systematic error relating to 
the momentum, chiral, and possibly also continuum extrapolations. 
Here we show that the uncertainty surrounding the 
momentum extrapolation can be eliminated with flavor twisted 
boundary conditions. With these imposed, one can extract the 
charge radius from simulations at zero Fourier momentum.

As boundary conditions modify the long-range physics on the lattice,
we must be careful to address effects of the finite volume.
In~\cite{Sachrajda:2004mi} volume effects for single-particle matrix elements
with twisted boundary conditions were shown to be similar to those with periodic
boundary conditions. Thus, provided one remains below multi-particle
cuts, the effect of the finite volume is exponentially suppressed in asymptotic volumes.  
In practice, lattice volumes are not asymptotically large, and the study with partially 
twisted boundary conditions in~\cite{Tiburzi:2006px} found sizable volume corrections 
to the nucleon isovector magnetic moment, $\sim 25 \%$, even for $m_\pi L \approx 4$. 
The bulk of such corrections arises from finite volume terms 
that do not respect hypercubic invariance. Physically one expects
contributions of this type because twisted boundary conditions 
introduce a particular direction in space. 
In this work, we determine
hypercubic breaking corrections to the pion dispersion relation
for partially twisted boundary conditions.
These corrections result in a 
finite volume modification to the pion's momentum.
Theoretical control over these volume 
effects is necessary wherever twisted boundary conditions are employed
because of the dominance of long-range pion physics in low-energy hadronic observables. 
Additionally we determine the finite volume effects for the  
extraction of the pion charge radius from lattice simulations with 
partially twisted boundary conditions. We further demonstrate numerically 
that such corrections are on the order of a few percent on current dynamical lattices.

Our presentation has the following organization.
In Section~\ref{disp}, we briefly review partially 
twisted boundary conditions and their incorporation
into partially quenched chiral perturbation theory (PQ\CPT). 
By calculating the pion self-energy, we determine the 
form of the pion dispersion relation in a finite box
and demonstrate that the induced momentum due to twisted
boundary conditions is modified by volume effects. 
The isopsin splitting among the pions is also evaluated in PQ\CPT. 
Next in Section~\ref{emff}, we show that the pion electromagnetic
form factor can be determined on the lattice at continuous 
values of the momentum transfer. Using PQ\CPT, we then 
determine the finite volume corrections resulting from partially
twisted boundary conditions. These volume effects are a 
controlled theoretical error that must be removed in order
to isolate the pion charge radius from lattice data. 
A summary (Section~\ref{summy}) concludes the paper.

\section{Dispersion relation with twisted boundary conditions} \label{disp}

The quark part of the partially twisted, partially quenched
QCD Lagrangian reads
\begin{equation} \label{eq:Ltwist}
\cL 
= 
\ol{Q}
\left(
\Dtslash + m_Q \right)
Q
,\end{equation}
where the six quark fields have been accommodated in the vector
$Q = (u, d, j, l, \tilde{u}, \tilde{d})^{\text{T}}$
that transforms in the fundamental representation of the
graded group $SU(4|2)$. The mass matrix in the isospin 
limit of the valence and sea sectors is given by
$m_Q = \diag(m_u, m_u, m_j, m_j, m_u, m_u)$.
The valence $u$ and $d$ quarks are accompanied by equal mass bosonic ghost quarks $\tilde{u}$ and
$\tilde{d}$. Functional integration over both valence and ghost quarks thus produces 
no net determinant factor, while integration over the sea quarks $j$ and $l$ produces a 
determinant.  The graded symmetry of $SU(4|2)$ hence provides a way
to write down a theory corresponding to partially quenched QCD. 
For a thorough discussion, see~\cite{Sharpe:2001fh}.

All $Q$ fields appearing in the Lagrangian 
satisfy periodic boundary conditions, 
and the effect of partial twisting has the form of a uniform
$U(1)$ 
gauge field:  
$\Dt_\mu = D_\mu + i B_\mu$, 
with the flavor matrix
$B_\mu = \diag (B^u_\mu, B^d_\mu, 0, 0, B^u_\mu, B^d_\mu )$, see~\cite{Sachrajda:2004mi,Tiburzi:2005hg} for details. 
The field 
$B_\mu$ 
acts as a flavor dependent field momentum:
for the 
$u$ 
quarks 
$B^u_\mu = ( 0,  \bm{\th}^u / L)$, 
and similarly for 
$d$.
The twist angles 
$\bm{\th}$ 
can be varied continuously 
and cannot be renormalized by short distance physics.
Notice that the twists are not implemented in the sea sector of the theory; thus, 
new gauge configurations do not need to be generated for each value of $\bm{\theta}$.

The dynamical effects due to partial twisting can be addressed using the low-energy effective
theory of partially quenched QCD in a finite volume. This low-energy theory is \PQCPT\
which describes the dynamics of the Goldstone bosons that arise from spontaneous
breaking of chiral symmetry. 
We determine volume effects in the $p$-regime of chiral perturbation theory~%
\cite{Gasser:1987zq} in which the longest-range fluctuations of the Goldstone modes 
do not restore chiral symmetry in a finite volume.  
The effective theory \PQCPT\ is written in terms of a periodic coset field 
$\S = \exp ( 2 i \phi / f)$ into which the meson fields $\phi$ are
embedded non-linearly. 
To incorporate partially twisted boundary conditions, we 
couple the field $\S$ to the background gauge field 
$B_\mu$. In terms of the field $\S$, 
the Lagrangian of \PQCPT\ appears as~%
\cite{Bernard:1994sv,Sharpe:1997by,Golterman:1998st,Sharpe:2000bc,Sharpe:2001fh}
\begin{equation} \label{eq:L}
\cL = 
\frac{f^2}{8} \str \left( \Dt_\mu \S \Dt_\mu \S^\dagger \right)
- 
\l \, \str \left( m_Q^\dagger \S + \S^\dagger m_Q \right)
+ 
\mu_0^2 \Phi_0^2
.\end{equation}
With 
$\mu_0 \to \infty$, 
the flavor singlet field 
$\Phi_0 = \str (\Sigma) / \sqrt{2}$
is integrated out, but the resulting flavor neutral meson propagators have both 
flavor connected and disconnected (hairpin) contributions~\cite{Sharpe:2001fh}. 
The action of the covariant derivative 
$\Dt^\mu$ 
is specified by~\cite{Sachrajda:2004mi}
\begin{equation} \label{eq:mmom}
\Dt_\mu \S 
= 
\partial_\mu \S 
+ 
i [ B_\mu, \S ] 
+ 
i A_\mu^a \left[ \ol T {}^a, \S \right]
,\end{equation}
where we have additionally included an isovector source field $A^a_\mu$
which will be utilized below in Section~\ref{emff}. 
The block diagonal form of the super-isospin matrices 
$\ol T {}^a$ is chosen to be $\ol T {}^a = (T^a , 0, T^a )$,
where $T^a$ are the usual isospin generators~\cite{Tiburzi:2006px}. 
From the commutator structure that couples the background gauge field $B_\mu$
to $\Sigma$, one sees that mesons have a flavor-dependent momentum boost~\cite{Sachrajda:2004mi}.
Expanding $\cL$ to tree level, one finds that mesons with 
quark content $Q \ol Q {}^\prime$ have masses\footnote{%
We shall often write $m_\pi$, and $m_{\pi^0}$ to 
denote the mesons formed from purely valence quarks. 
Other cases for which we also use $m_\pi$ as an abbreviation
will be clarified in the text.
}
\begin{equation}
m_{QQ'}^2 = \frac{\lambda}{4 f^2} ( m_Q + m_{Q'})
,\end{equation}
and kinematic momenta
\begin{equation} \label{eq:momenta}
\bm{P}_{Q Q'} = \bm{p} + \bm{B}_Q - \bm{B}_{Q'} 
,\end{equation} 
where $\bm{p}$ is the Fourier three-momentum.

\begin{figure}[tb]
  \includegraphics[width=0.4\textwidth]{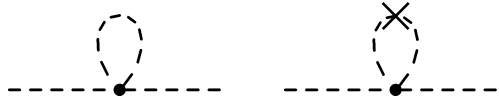}%
  \caption{
    	Pion self-energy diagrams in \PQCPT.  Dashed lines represent mesons, 
        while the cross represents the partially quenched hairpin interaction.  
  }
  \label{f:twpion}
\end{figure}

Beyond tree-level, observables receive non-analytic 
corrections from loop diagrams generated in the effective theory.
These loop diagrams, moreover, are modified in finite volume.
As twisted boundary conditions are flavor dependent, isospin 
symmetry is broken and the degeneracy among 
the pions is lifted by volume effects.
The isospin splitting among the pions 
was determined for completely twisted \CPT\ in~%
\cite{Sachrajda:2004mi}. Here we determine the splitting in
partially twisted \PQCPT. Moreover as
twisted boundary conditions introduce a particular direction
in space, the pion dispersion relation is no longer
a hypercubic invariant function of the Fourier momentum $\bm{p}$. 
We determine the form of the dispersion relation and show that
the kinematic momenta in Eq.~\eqref{eq:momenta} are consequently modified.

To determine these volume effects, we must calculate 
the one-loop diagrams contributing to the self-energy of
the pions. These diagrams are depicted in Fig.~\ref{f:twpion}.
Evaluating the contractions leading to the diagrams in the Figure,
we are confronted with various terms that can be grouped according
to where the two derivatives of the four-pion vertex act. 
Let us detail the calculation of the diagrams for the charged pion.
Terms where both derivatives at the vertex act on the internal line
or both act on the external legs are the most familiar because these 
survive the infinite volume limit. The former give rise to a mass shift, 
and the latter constitute a wavefunction renormalization correction.
A general contribution to a loop diagram in finite volume, $C_L$, can be written as
\begin{equation}
C_L = C_\infty + \delta_L(C)
,\end{equation}
where $C_\infty$ is the result for contribution $C$ in infinite volume
and the difference $\delta_L (C) = C_L - C_\infty$ is ultraviolet finite.
Combining the mass shift and contributions from wavefunction renormalization,
we arrive at $\delta_L(m_{\pi^\pm}^2)$ the finite volume shift of the 
charged pion mass which is given by
\begin{equation} \label{eqn:chargedpionmassfinitevolume}
\d_L \left( m_{\pi^\pm}^2 \right) 
=
\frac{m_\pi^2}{2 f^2} 
\frac{\partial}{\partial m_\pi^2}
\left[
(m_\pi^2 - m_{jj}^2) 
\cI_{1/2}(\bm{0}, m_\pi^2)
\right]
.\end{equation} 
As is well known, due to delicate cancellations the charged pion mass
only receives loop contributions from flavor disconnected (hairpin) propagators. 
Thus so too must the finite volume correction.
These hairpin contributions are most economically written in terms
of derivatives with respect to the pion mass squared, as appears 
in Eq.~\eqref{eqn:chargedpionmassfinitevolume}. Because the net
contribution to the mass is from flavor neutral loop mesons, there 
is no $\bm{B}$-dependence in the volume effect at this order. 
The function $\cI_{\beta}(\bm{B},m^2)$  results from taking the 
difference of the finite volume contribution minus the infinite volume contribution, 
and is given by
\begin{eqnarray} \label{Ifunc}
\cI_{\b}(\bm{B},m^2) &=& \frac{1}{L^3} \sum_{\bm{k}} 
\frac{1}{[(\bm{k} + \bm{B})^2 + m^2]^\b}
- 
\int
\frac{d\bm{k}}{(2 \pi)^3}
\frac{1}{[(\bm{k}+ \bm{B})^2 + m^2]^\b} 
.\end{eqnarray} 
To evaluate this finite volume function, the sum over
Fourier momentum modes can be cast into an exponentially 
convergent form by using the Poisson re-summation formula. 
Resulting expressions are then swiftly evaluated using
Jacobi elliptic functions, see~\cite{Zinn-Justin:2002ru,Sachrajda:2004mi,Bunton:2006va}.

We are not done evaluating the momentum contractions for the diagrams
shown in the Figure. There are also contributions where one derivative 
acts on the internal line and one derivative acts on either of the external legs. 
These contractions yield terms proportional to
\begin{equation} \label{newsum}
\frac{1}{L^3} 
\sum_{\bm{k}} 
\frac{(\bm{p} + \bm{B}_{\pi^\pm}) \cdot (\bm{k} + \bm{B}_q)}{\sqrt{(\bm{k} + \bm{B}_q)^2 + m_{jq}^2}} 
,\end{equation}
where $q$ is a valence quark index and $m_{jq}^2$ is the mass of the corresponding valence-sea meson
formed from $q$ and either of the degenerate sea quarks. As such contractions involve only
valence-sea meson loops, they are absent the quenched approximation. 
Notice that in the infinite volume limit, the sum in Eq.~\eqref{newsum} can be converted into an integral.
Using a shift of variables, the infinite volume contribution is proportional to 
\begin{equation} \notag
\int \frac{d\bm{k}}{(2\pi)^3}  \, \frac{(\bm{p} + \bm{B}_{\pi^\pm}) \cdot \bm{k}}{\sqrt{\bm{k}^2 + m_{jq}^2}} 
,\end{equation}
and vanishes by $SO(3)$ invariance. 
Furthermore in a lattice with periodic boundary conditions ($\bm{B}_u = \bm{B}_d = \bm{0}$) the sum in
Eq.~\eqref{newsum} becomes
\begin{equation} \notag
\frac{1}{L^3}
\sum_{\bm{k}} \frac{\bm{p} \cdot \bm{k}}{\sqrt{\bm{k}^2 + m_{jq}^2}} 
,\end{equation}
and vanishes by cubic invariance. With twisted boundary conditions, however,
the sum in Eq.~\eqref{newsum} does not vanish. Hence terms that ordinarily vanish
in infinite volume due to rotational invariance (and similarly vanish in a periodic box due to 
cubic invariance) contribute. The sum, moreover, is non-vanishing only in the directions
for which $\bm{B}_q$ is non-vanishing. These contributions then are clearly due to 
the breaking of cubic invariance introduced by the spatial direction specified by the background gauge field. 
To describe finite volume effects from these terms, we define the
function
\begin{equation} \label{Kfunc}
\bm{\mathcal{K}}_{\b} (\bm{B}, m^2)
=
\frac{1}{L^3} \sum_{\bm{k}} 
\frac{k^j + B^j}{[(\bm{k} + \bm{B})^2 + m^2]^\b}
,\end{equation}
which can be evaluated in terms of a derivative of the elliptic Jacobi function~\cite{Tiburzi:2006px}.

Combining all effects due to the finite volume, 
we find that the charged pion dispersion relation has the form\footnote{%
In order to arrive at Eq.~\eqref{eq:pionenergy}, we added a higher-order term 
proportional to $\bm{K}^2$ which scales as $f^{-4}$. The one-loop results
scale as $f^{-2}$. 
}
\begin{equation} \label{eq:pionenergy}
E_{\pi^\pm}(\bm{p})^2 
=
(\bm{p} + \bm{B}_{\pi^\pm} \pm \bm{K})^2 + m_\pi^2 + \d_L \left( m_{\pi^\pm}^2 \right) 
,\end{equation}
where the function $\bm{K}$ 
appears as an additive renormalization to the field momentum 
$\bm{B}_{\pi^\pm} = \pm (\bm{B}_u - \bm{B}_d)$, 
and is given by
\begin{equation} \label{eq:K}
\bm{K} = - \frac{1}{f^2}[ \bm{\mathcal{K}}_{1/2} (\bm{B}_u, m_{ju}^2)-  \bm{\mathcal{K}}_{1/2} (\bm{B}_d, m_{ju}^2) ]
.\end{equation} 
From Eq.~\eqref{eq:K}, we see that the pion field momentum is renormalized only in the directions
for which $\bm{B}_{\pi^\pm}$ is non-vanishing. 
Furthermore, the analogous quenched finite volume 
contribution vanishes.

\begin{figure}[tb]
	\bigskip		
  \includegraphics[width=0.45\textwidth]{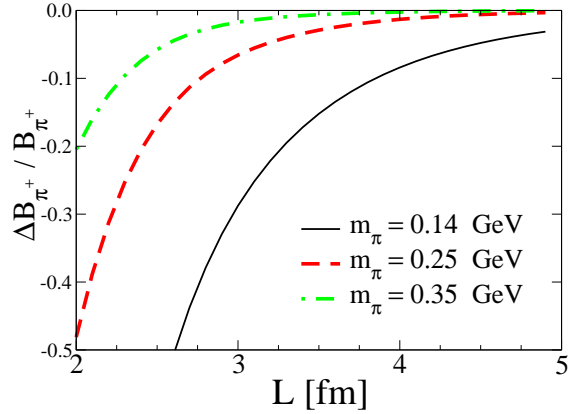}%
  \caption{
    	Maximal change in the induced momentum due to twisting.   
        Here, the pion mass $m_\pi$ is the valence-sea meson mass, 
        and $\bm{\theta}^u = - \bm{\theta}^d$ in the limit of vanishing 
        twist. 
  }
  \label{f:renB}
\end{figure}

\begin{figure}[tb]
  \includegraphics[width=0.45\textwidth]{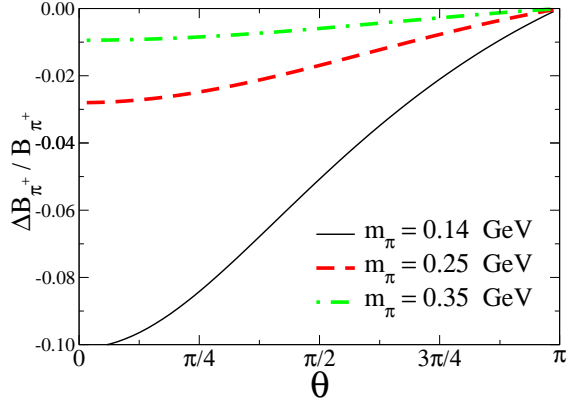}%
  \caption{
    	Relative change in the induced momentum due to twisting as a function of the twist angle.   
        The pion mass $m_\pi$ refers to the valence-sea meson mass, and the box size
        is fixed at $2.5 \, \texttt{fm}$.
  }
  \label{f:renB2}
\end{figure}

We can investigate the finite volume renormalization of the 
field momentum by considering the relative change in the 
pion's momentum due to twisting
\begin{equation} \label{eq:renB}
\D B_{\pi^+} / B_{\pi^+} = \sum_{j} K^j / B_{\pi^+}^j
.\end{equation}
The relative change in the negatively charged pion's momentum is numerically
identical. In the limit 
$\bm{B}_{\pi^+} \to \bm{0}$, 
we have 
$\D B_{\pi^+} / B_{\pi^+} = \bm{\nabla}_B \cdot \bm{K}$.
One should keep in mind, however, that neither of these quantities
is accessible without twisted boundary conditions, but the twisted
divergence serves as an approximate relative change for small twists.
In Fig.~\ref{f:renB}, we plot the relative change in induced momentum
$\D B_{\pi^+} /  B_{\pi^+}$ in Eq.~\eqref{eq:renB} for the worst case scenario. 
We take $\bm{\theta}^u = - \bm{\theta}^d$ and further choose each component 
near zero. 
Finite volume modifications for this case 
are seen to be substantial from the figure. 
Thus in applications of twisted boundary conditions, 
one must take care to avoid large additive volume renormalization
of the twisting parameters. Choosing only one flavor to be 
twisted, e.g.~$\bm{\theta}^d = \bm{0}$, and implementing the 
twist in only one spatial direction reduces
the volume effect by a factor of six. 
Furthermore in the actual implementation
of twisted boundary conditions near vanishing twist angles
will not be employed. 
Thus in Fig.~\ref{f:renB2}, we investigate the 
behavior with respect to the single twist angle $\theta$, where 
$\bm{\theta}^u = (0,\theta,0)$ and $\bm{\theta}^d = \bm{0}$. 
The relative change in the induced momentum is plotted versus $\theta$
for a few valence-sea pion masses (denoted by $m_\pi$ in the figure), but with the 
box size $L = 2.5 \, \texttt{fm}$ fixed.
The behavior with $\theta$ is damped oscillatory and crosses zero
when $\theta = (2 n + 1) \pi$.
For reasonably small values of the twist angle, $\theta \sim \pi / 8$,
and at (valence-sea) pion masses $\sim 0.3 \, \texttt{GeV}$,  
the induced momentum is shifted by $\sim 2\%$ from volume effects.
Thus the finite volume renormalization of the twist angles can practically be ignored
on current lattices.
As pion masses are brought down, however, 
volume renormalization of the induced momentum 
may need to be accounted for by evaluating Eq.~\eqref{eq:K}.

The above discussion applies to the charged pions $\pi^\pm$. 
The neutral pion dispersion relation, however, is unaffected by twisting.
Momentum contractions where one derivative acts on the internal line and
one acts on the final state are canceled by those where the second
derivative acts on the initial state.  
Thus only the mass of the neutral pion is affected by the finite volume.
Performing the self-energy calculation for the $\pi^0$, we find the finite volume mass shift
\begin{equation} \label{eqn:neutralpionmassfinitevolume}
\d_L \left( m_{\pi^0}^2 \right) 
=
\frac{m_\pi^2}{2 f^2} 
\left\{
\frac{\partial}{\partial m_\pi^2}
\left[
(m_\pi^2 - m_{jj}^2) 
\cI_{1/2}(\bm{0}, m_\pi^2)
\right]
+ 
2 \, \cI_{1/2}(\bm{B}_{\pi^+},m_\pi^2) 
-
2 \, \cI_{1/2}(\bm{0},m_\pi^2) 
\right\}
.\end{equation}

\begin{figure}[tb]
\bigskip
  \includegraphics[width=0.45\textwidth]{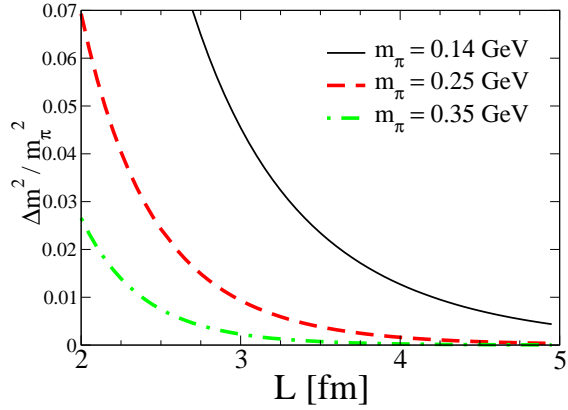}%
  \caption{
    	Maximal isospin splitting between the charged and neutral pions.   
  }
  \label{f:split}
\end{figure}

Combining the finite volume shifts for the charged and neutral 
pions, we find that the isospin splitting due to partially twisted boundary conditions
has the form
\begin{eqnarray} \label{eq:split}
\D m^2 
&=& 
\d_L 
\left( 
m_{\pi^\pm}^2 
\right) 
- 
\d_L 
\left(
m_{\pi^0}^2
\right) 
\notag \\
&=& 
\frac{m_\pi^2}{f^2}
[
\cI_{1/2}(\bm{0},m_\pi^2) 
- 
\cI_{1/2}(\bm{B}_{\pi^+},m_\pi^2) 
]
,\end{eqnarray}
where the hairpin contributions have canceled out. 
The splitting $\D m^2$ is only sensitive to the valence pion mass
and vanishes when the twists preserve isospin, 
$\bm{B}_u = \bm{B}_d$.
In Fig.~\ref{f:split}, we plot the relative isospin splitting
$\D m^2 / m_\pi^2$ 
as a function of the lattice size 
$L$ 
for a few values of the valence pion mass 
$m_\pi$. 
The maximal isospin splitting occurs for 
$\bm{\theta}_{\pi^+} = ( \pi, \pi, \pi)$, 
and we have chosen to plot this value to investigate the worst case scenario. 
From the figure, we see $\sim 10\%$ maximal isospin 
splitting at the physical pion mass in a relatively small box. 
For lattice pion masses $\sim 0.3 \, \texttt{GeV}$, the maximal 
isospin splitting is $\sim 2 \%$ in a $2.5 \, \texttt{fm}$ box. 
For applications of twisted boundary conditions, such as 
the extraction of the pion charge radius, only one component
of $\bm{\theta}_{\pi^+}$ need be non-vanishing. Moreover, 
this twist angle should be chosen less than $\pi$ to reach
smaller momentum transfer.
These considerations restrict the isospin splitting 
to be $< 1 \%$ for pion masses $\sim 0.3 \, \texttt{GeV}$ in a $2.5 \, \texttt{fm}$
box. For this reason we neglect the isospin splitting in virtual loop meson masses below.

Having detailed the finite volume corrections to the pion 
mass and dispersion relation, we now describe how partially
twisted boundary conditions enable the extraction of the
pion electromagnetic form factor.

\section{Pion electromagnetic form factor} \label{emff}

\begin{figure}[tb]
  \centering
  \includegraphics[width=0.25\textwidth]{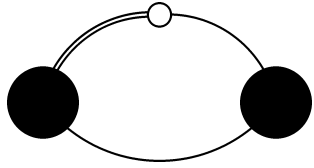}%
  \caption{Contraction for the pion form factor in Eq.~\eqref{eq:pionkey}.
  The source and sink are denoted by filled circles, the operator insertion by an 
  open circle. Single lines represent twisted $d$-quark propagators, 
  while the double line represents the twisted $u$-quark propagator. 
  }
  \label{f:contract}
\end{figure}

The pion electromagnetic form factor $G_\pi(q^2)$ arises in the charged 
pion matrix element of the electromagnetic current $J_\mu^{\text{em}}$, namely
\begin{equation}
\langle \pi^+ (\bm{p}')  | \, J_\mu^{\text{em}} \, | \pi^+ (\bm{p}) \rangle 
=
(p' + p)_\mu G_\pi (q^2)
,\end{equation} 
with 
$q = p' - p$.
In the isospin limit, however, vector flavor symmetry
relates this matrix element to the isospin transition~%
\cite{Tiburzi:2006px}
\begin{equation} \label{eq:pionkey}
\langle \pi^+ (\bm{p}')| \, \ol u \, \gamma_\mu \, d  | \, \pi^0 (\bm{p})  \rangle
= 
- \sqrt{2} \, \langle \pi^+ (\bm{p}') | \, J_\mu^{\text{em}} \, | \pi^+  (\bm{p})\rangle 
,\end{equation}
where we have made use of charge conjugation invariance, and the
phase convention is that commonly employed in \CPT.
Thus by calculating the pion form factor from the left-hand side
of the above equation, we can induce momentum transfer with partially
twisted boundary conditions~%
\cite{Tiburzi:2005hg}.
The neutral pion is flavor non-singlet and because we work in the 
isospin limit, the annihilation diagrams contributing to the left-hand
side of Eq.~\eqref{eq:pionkey} exactly cancel. Thus only 
the quark contraction shown in Fig.~\ref{f:contract} must be calculated in lattice simulations. 
While flavor neutral states in 
partially twisted (and partially quenched) theories
have sicknesses related to lack of unitarity~%
\cite{Sharpe:2000bc,Sharpe:2001fh}, 
the illness does not infect the states with isopsin $I = 1$. 
Consequently the propagator of the $\pi^0$ is diagonal 
with a simple pole form, and asymptotic $\pi^0$ states 
can be created \`a la LSZ.

Application of partially twisted boundary conditions
to the matrix element in Eq.~\eqref{eq:pionkey} thus leads 
to the kinematical effect
\begin{equation} \label{eq:novolume}
\langle \pi^+ (\bm{0}) | \, \ol u \, \gamma_4 \, d  | \, \pi^0 (\bm{0}) \rangle
= 
- \sqrt{2} ( p'_4 + p_4 ) G_\pi (B^2)
,\end{equation}
where for emphasis we have projected the initial and final states onto zero Fourier momentum,
and restricted our attention to the fourth component of the current.
The energy of the initial state is 
$p_4 = i m_\pi$ 
and the energy of the final state is
$p'_4 = i \sqrt{m_\pi^2 + \bm{B}_{\pi^+}^2}$, 
while the momentum transfer squared is 
$B^2 = 2 m_\pi^2 \left( - 1 + \sqrt{1 + \bm{B}_{\pi^+}^2 / m_\pi^2}\right)$. 
Finally the induced three-momentum transfer 
$\bm{B}_{\pi^+}$ 
is given by the difference of twists of the flavors changed 
$\bm{B}_{\pi^+} = \bm{B}_u - \bm{B}_d$. 
In writing the above expression, we have temporarily ignored dynamical effects
from the finite volume. 
To assess the effects of the finite volume on the pion electromagnetic
form factor with partially twisted boundary conditions, we must again use PQ\CPT\ and do so in the $p$-regime.

\begin{figure}[tb]
  \includegraphics[width=0.4\textwidth]{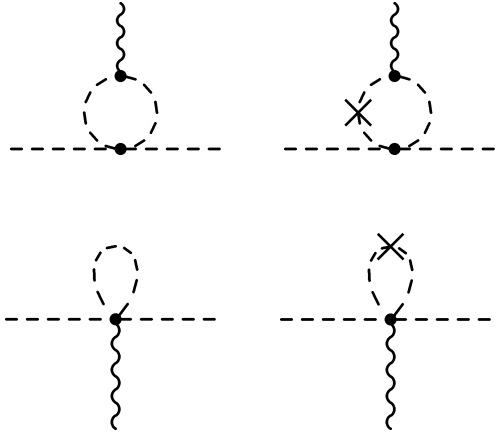}%
  \caption{
    One-loop contributions to the isospin transition matrix element
    in \PQCPT. The wiggly line shows the insertion of the isospin raising 
    operator. Other diagram elements are the same as in Fig.~\ref{f:twpion}.
  }
  \label{f:twpionFF}
\end{figure}
At next-to-leading order in the EFT, there are local contributions from higher-order
operators and loop contributions generated from the leading-order vertices. We handle 
the local contributions first.
In partially twisted \PQCPT, the isovector current operator $J_\mu^+$ 
receives a local contribution at next-to-leading order from the term
\begin{equation} \notag
\d J^+_\mu 
= 
i \alpha_9 \,
\Dt_\nu \,
\str
\left[ 
\ol T {}^+
\left( 
[ \Dt_\mu \S^\dagger, \Dt_\nu \S ]
-
[\Dt_\nu \S^\dagger, \Dt_\mu \S ]
\right)
\right]
.\end{equation}
This is just a rewriting of the current derived from the $\alpha_9$ term of the Gasser-Leutwyler 
Lagrangian~\cite{Gasser:1983yg}.
As explained in~\cite{Tiburzi:2006px}, the derivatives appearing in operators for external currents must
be upgraded to $\tilde{D}$'s.
There are also local contributions to the current from the $\a_4$ and $\a_5$ terms of the fourth-order 
Lagrangian, but these are canceled by wavefunction renormalization, see~\cite{Gasser:1984ux,Donoghue:1992dd}.

The one-loop contributions to the isovector form factor are shown in Fig.~\ref{f:twpionFF}
(along with the requisite wavefunction corrections shown in Fig.~\ref{f:twpion})
and lead to finite volume modifications. 
The hairpin diagrams exactly cancel in both infinite and finite volumes. 
This can be understood easily due to the isospin relation in Eq.~\eqref{eq:pionkey}. 
On the right-hand side of this relation is the matrix element of the electromagnetic current.
Contributions to this matrix element at one-loop are only due to charged meson loops, 
thus there can be no hairpins which arise from flavor (and thus charge) neutral mesons. 
Explicit calculation of the left-hand side of Eq.~\eqref{eq:pionkey} verifies this.

Returning to Eq.~\eqref{eq:novolume}
but now including the effect of the finite volume at one-loop order in the
chiral expansion, we have
\begin{align} \label{eq:result}
\langle \pi^+ (\bm{p}') | \, \ol u \, \gamma_4 \, d  | \, \pi^0 (\bm{p}) \rangle
= 
- \sqrt{2} i 
\Bigg\{
& [ E_{\pi^+}(\bm{p}') + E_{\pi^0}(\bm{p}) ]
\left[ 
G_\pi (Q^2) + G_{FV} 
\right]
\notag\\
+ 
& [ E_{\pi^+}(\bm{p}') - E_{\pi^0}(\bm{p}) ]
\left[
G_{FV}^{\text{iso}} + (\bm{p}' + \bm{B}_{\pi^+} + \bm{p}) \cdot \bm{G}_{FV}^{\text{rot}}
\right]
\Bigg\}
,\end{align}
where
$E_{\pi^+}(\bm{p}')$ is given in Eq.~\eqref{eq:pionenergy}, and
$E_{\pi^0}(\bm{p})^2 = \bm{p}^2 + m_\pi^2 + \d_L (m_{\pi^0}^2 )$. Above $Q$ is the
effective momentum transfer, $Q = q + B_{\pi^+}$, 
and $G_\pi(Q^2)$ is the partially quenched pion electromagnetic form factor in the infinite 
volume limit~\cite{Arndt:2003ww}, namely
\begin{equation} \label{eq:FF}
G_\pi(Q^2) 
= 
1 - \frac{4 \alpha_9}{f^2} Q^2 
+ 
\frac{1}{48 \pi^2 f^2} 
\left[ 
Q^2 \log \frac{m_{ju}^2}{\mu^2} 
+ 
4 m_{ju}^2 \mathcal{F} \left( \frac{-Q^2}{4 m_{ju}^2} \right)
\right] 
,\end{equation} 
where the auxiliary function $\mathcal{F}(x)$ is defined by
\begin{equation}
\mathcal{F}(x) = (x - 1) \sqrt{1 - \frac{1}{x}} \log \frac{\sqrt{1 - \frac{1}{x} + i \varepsilon}-1}
{\sqrt{1 - \frac{1}{x} + i \varepsilon} + 1} + \frac{5}{3} x - 2
.\end{equation}
The remaining contributions in Eq.~\eqref{eq:result} are finite volume modifications
and have been denoted so with a $FV$ subscript. There are various flavor and 
momentum contractions for the meson fields that lead to these terms. They fall into 
three categories. We give the explicit results for each, and comment on the physics behind such terms.

The first finite volume contribution $G_{FV}$ is the finite volume modification to 
the form factor. This contribution arises from the difference between 
the sum of Fourier modes in the loop and the infinite volume loop integral that leads to Eq.~\eqref{eq:FF}. 
This effect remains even in the absence of twisted boundary 
conditions. Explicitly we have
\begin{align} 
G_{FV} = 
\frac{1}{2 f^2} \int_0^1 dx 
\Bigg[ &
\cI_{1/2} 
\left(
x\bm{Q} - \bm{B}_u,m_{ju}^2 + x(1-x) Q^2
\right) 
- 
\cI_{1/2} (-\bm{B}_u, m_{ju}^2 ) 
\notag \\
& + 
\cI_{1/2} 
\left(
x\bm{Q} + \bm{B}_d,m_{ju}^2 + x(1-x) Q^2
\right) 
- 
\cI_{1/2} (\bm{B}_d, m_{ju}^2 ) 
\Bigg] \label{eq:GFV}
,\end{align}
and when 
$\bm{B}_u = \bm{B}_d = \bm{0}$, 
we recover the finite volume
modification to the form factor with periodic boundary conditions~\cite{Bunton:2006va}.

The remaining finite volume effects in Eq.~\eqref{eq:result} are due to symmetry breaking introduced
by the partially twisted boundary conditions. 
The second finite volume contribution 
$G_{FV}^{\text{iso}}$ 
arises from the breaking of isospin and vanishes when 
$\bm{B}_u = \bm{B}_d$.
This term moreover enters the matrix
element in Eq.~\eqref{eq:result} proportional to the energy difference between the initial and
final states. Ordinarily such an additional structure is forbidden 
by current conservation. 
In finite volume, however, the partially twisted boundary conditions
break isospin symmetry and the isospin current $J_\mu^+$ is no longer conserved. 
This finite volume correction 
$G_{FV}^{\text{iso}}$ 
appears as
\begin{align}
G_{FV}^{\text{iso}}
= 
\frac{1}{6 f^2} 
\int_0^1 dx (1 - 2 x) 
& \Bigg\{
4 \, \cI_{1/2}
\left(
x \bm{Q}, m_\pi^2 + x(1-x)Q^2
\right)
\notag \\
- &
Q^2 \left[
1 + 2  x(1-x)  
\right]
\cI_{3/2}
\left( 
x \bm{Q}, m_\pi^2 + x(1-x) Q^2
\right)
\notag \\
+
& (1- 2 x) \bm{Q} \cdot \bm{\mathcal{K}}_{3/2}
\left( x \bm{Q}, m_\pi^2 + x(1-x)Q^2
\right)
\Bigg\} -\frac{1}{3} \D m^2 / m_\pi^2
,\end{align}
and involves the isospin splitting among the pions $\D m^2$ given in Eq.~\eqref{eq:split}.

The final finite volume modification $\bm{G}_{FV}^{\text{rot}}$ arises from 
the breaking of hypercubic invariance introduced by partially twisted boundary conditions.
This contribution is given by
\begin{align}
\bm{G}_{FV}^{\text{rot}}
= 
\frac{1}{4 f^2}
\int_0^1 dx (1- 2x) 
& \Bigg[
\bm{\mathcal{K}}_{3/2} 
\left(
x \bm{Q} - \bm{B}_u, m_{ju}^2 + x (1-x) Q^2  
\right)
\notag \\
& + 
\bm{\mathcal{K}}_{3/2} 
\left(
x \bm{Q} + \bm{B}_d, m_{ju}^2 + x (1-x) Q^2  
\right)
\Bigg]
.\end{align}
Notice this term only vanishes when $\bm{B}_u = \bm{B}_d = 0$.
When $\bm{B}_u = \bm{B}_d \neq 0$, isospin symmetry is restored
and we might expect $\bm{G}_{FV}^{\text{rot}}$ to vanish given 
the way this term enters Eq.~\eqref{eq:result}. 
This is not the case, however, because twisted boundary 
conditions break cubic invariance and extra terms are allowed.
Indeed the decomposition of the current matrix element
in the case when the twisted boundary conditions preserve 
isospin $\bm{B}_u = \bm{B}_d \neq 0$,
\begin{equation} \label{eq:mystery}
\langle \pi^+ (\bm{p}') | \, J^+_\mu \,  | \, \pi^0 (\bm{p}) \rangle
= 
(p' + p)_\mu G(Q^2)
,\end{equation}
relies on $SO(4)$ rotational invariance. While this 
reduces to hypercubic invariance on the lattice, if there are no terms in the
EFT that violate $SO(4)$ at this order, then Eq.~\eqref{eq:mystery} still holds. 
Twisted boundary condition, however, do not respect the hypercubic invariance of the lattice
and the above form is no longer valid.  
Explicitly $\bm{G}_{FV}^{\text{rot}}$ is not invariant under the 
transformation $Q_j \to - Q_j$, and indeed breaks hypercubic invariance
even in the isospin symmetric case. We should thus interpret this term in Eq.~\eqref{eq:result}
as arising from cubic symmetry breaking rather than the violation of isospin.\footnote{%
As for the other finite volume contributions, $G_{FV}$ in Eq.~\eqref{eq:GFV} is not invariant
under $Q_j \to - Q_j$ unless $\bm{B}_u = \bm{B}_d$. 
The contribution $G_{FV}^{\text{iso}}$ is invariant under $Q_j \to - Q_j$
and so its presence in Eq.~\eqref{eq:result} is unambiguously due to isospin violation.
}

\begin{figure}[tb]
\bigskip
  \includegraphics[width=0.45\textwidth]{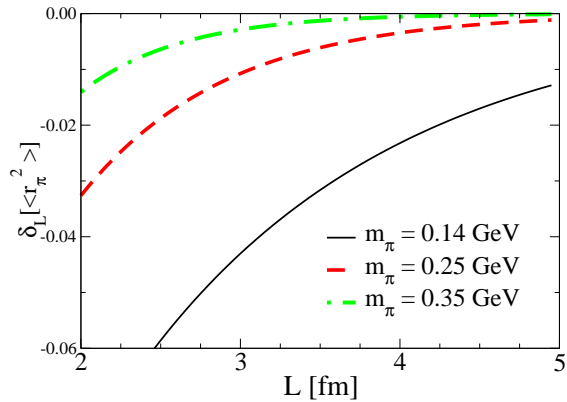}%
  \caption{
       Finite volume shift of the pion charge radius with partially twisted 
       boundary conditions. Plotted versus the box size $L$ is the
       relative change in the charge radius $\d_L[<r_\pi^2>]$ for a few values of the 
       pion mass. 
  }
  \label{f:radius}
\end{figure}

Having deduced the volume corrections to the pion form factor using PQ\CPT, 
we can now investigate the impact of partially twisted boundary conditions
on the extraction of the pion charge radius. To do this, we choose 
for simplicity $\bm{B}^d = \bm{0}$ and $\bm{B}^u = (0, |\bm{B}|, 0)$, with $|\bm{B}| = \theta / L$. 
Furthermore we take the source and sink to be projected onto zero 
Fourier momentum.
We investigate the relative difference of the form factor (minus the isovector charge)
in finite volume to infinite volume. Specifically we investigate the following quantity 
\begin{equation} \label{eq:FVdiff}
\delta_L [ G_\pi (Q^2) - 1] 
= 
\frac{\frac{1}{- \sqrt{2} (p'_4 + p_4)} 
\langle \pi^+ (\bm{0}) | J_4^+   | \, \pi^0 (\bm{0}) \rangle 
- 
G_\pi (Q^2)}{G_\pi(Q^2) - 1}
.\end{equation}
For small twist angles, $\theta \ll 1$, the relative difference in the
form factor $\delta_L [ G_\pi (Q^2) - 1]$ becomes the relative difference
in the charge radius
\begin{align}
\lim_{\theta \to 0} 
\delta_L [ G_\pi (Q^2) - 1] 
& \equiv
\delta_L [ < r_\pi^2 > ] 
\notag \\ 
& = \frac{<r_\pi^2>_L - <r_\pi^2>_\infty }{ <r_\pi^2>_\infty}
.\end{align} 
Finally as expressions depend on both the valence-valence and valence-sea meson masses, we
take the theory to be unquenched, so that $m_{ju}^2 = m_\pi^2$. 
In Fig.~\ref{f:radius}, we plot $\delta_L [ < r_\pi^2 > ]$
as a function of $L$ for various values of the pion mass to 
contrast our results with those of Ref.~\cite{Bunton:2006va}. 
We plot the total contribution which arises from the three finite volume terms in Eq.~\eqref{eq:result},
however, nearly all of the effect is due to $G_{FV}$. 
Contribution from terms that arise from the breaking of symmetries 
in finite volume, $G_{FV}^{\text{iso}}$ and $\bm{G}_{FV}^{\text{rot}}$, are suppressed
relative to $G_{FV}$
due to the kinematical pre-factor accompanying these terms. 
The magnitude of the finite volume effect shown in the figure is smaller than that 
calculated in~\cite{Bunton:2006va}, and has differing sign. 
In that study, however, periodic boundary conditions were assumed and 
the extraction of the charge radius required utilizing 
the smallest available Fourier momentum transfer.
With twisted boundary conditions, we have eliminated this restriction.
Volume effects are consequently different.

\begin{figure}[tb]
\bigskip
  \includegraphics[width=0.45\textwidth]{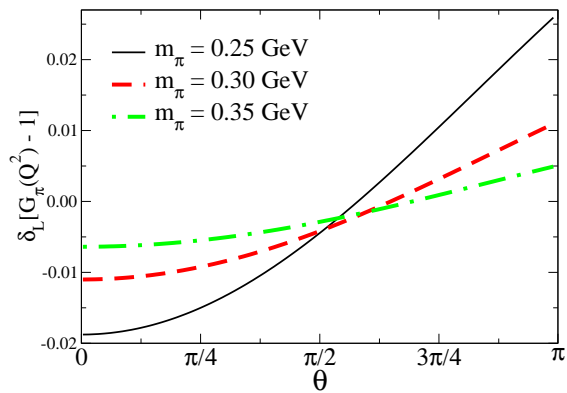}%
  \caption{
       Finite volume shift of the pion form factor with partially twisted 
       boundary conditions. Plotted versus $\theta$ is the 
       relative change in the form factor $\d_L[G_\pi(Q^2) -1]$ for a few values of the 
       pion mass. The three-momentum transfer from twisting is $|\bm{B}| = \theta \times 0.079 \, \texttt{GeV}$, 
and the box size is fixed at $2.5 \, \texttt{fm}$. 
  }
  \label{f:FVdiff}
\end{figure}

Of course to extract the charge radius from twisted boundary conditions, we require $\theta \neq 0$,
and thus we need to investigate $\delta_L [ G_\pi (Q^2) - 1]$ in Eq.~\eqref{eq:FVdiff}.
In Fig.~\ref{f:FVdiff}, we fix the lattice volume at $2.5 \, \texttt{fm}$
and plot the relative difference $\delta_L [ G_\pi (Q^2) - 1]$ as a function 
of $\theta$ for a few values of the pion mass. We see that the finite volume
effect increases with $\theta$ over the range plotted.
Generally the finite volume effects for form factors decrease with momentum transfer~\cite{Detmold:2005pt,Tiburzi:2006px}, 
because as the resolving power of the virtual probe increases,
the effect of the finite volume decreases. The behavior with increasing 
momentum transfer (here parametrized by $\theta$) is indeed damped oscillatory 
but the oscillations appear beyond the range of the EFT.
Nonetheless, the finite volume effect only amounts to a few percent
on current dynamical lattices.

Flavor twisted boundary conditions introduce systematic 
error into the determination of observables, such as the pion charge radius.
We stress, however, that this is a controlled effect. 
There are no extraneous phenomenological
form factor fitting functions needed. The EFT can be used for 
the momentum extrapolation because employing $\theta < \pi$ brings us into the
range of applicability of the EFT on current dynamical lattices.  Furthermore 
there are no additional parameters introduced in the EFT to describe
the volume dependence with twisted boundary conditions. For mesonic observables, 
knowledge of the pion decay constant $f_\pi$ and the pion mass $m_\pi$ (possibly 
also the valence-sea pion mass) is enough to remove the volume 
dependence. Higher-order dependence on the quark mass, volume 
and momentum transfer can all be calculated systematically in the EFT. 
In particular, the pion charge radius determined from twisted 
boundary conditions will be rather insensitive to volume effects
on current dynamical lattices, see Figs.~\ref{f:radius} and \ref{f:FVdiff}.

\section{Summary} \label{summy}

Above we have shown that in the isospin limit the pion electromagnetic form factor 
may be extracted from the isospin changing matrix element in Eq.~\eqref{eq:pionkey}.
Utilizing flavor twisted boundary conditions allows one to circumvent the restriction
to quantized momentum transfer. We have detailed the finite volume corrections to 
the pion mass and dispersion relation using partially twisted PQ\CPT. 
Addressing these effects is crucial to determining
observables, such as the pion charge radius, in simulations with twisted boundary conditions.

We find that the twist angles acquire additive finite volume renormalization which 
can be sizable near vanishing twist. This effect is only present for dynamical simulations,
and the renormalized twist angle $\bm{\theta}_q^{\text{ren}} (L)$, with the quark flavor $q = u$ or $d$, is given by
\begin{equation} 
\bm{\theta}_q^{\text{ren}}(L) 
= 
\bm{\theta}_q 
- 
\frac{L}{f^2} 
\bm{\mathcal{K}}_{1/2} 
\left( \frac{\bm{\theta}_q}{L}, m_{jq}^2 \right)
,\end{equation} 
where $j$ labels the sea quark (in a two-flavor degenerate sea). 
For practical applications of partially twisted boundary
conditions on current lattices this volume correction is $\sim 2\%$. 
In finite volume, partially twisted boundary conditions lead to isospin symmetry breaking. 
The volume induced isospin splitting among the pions is under
good theoretical control, $< 1 \%$ for practical applications of twisted
boundary conditions on current lattices. Finally, the extraction of the pion 
charge radius with partially twisted boundary conditions introduces no new 
free parameters, and finite volume effects can be assessed with effective field theory methods. 
We find that the finite volume effects for extracting the pion charge radius
from simulations with twisted boundary conditions are only a few percent.

A simple exercise in isopsin algebra~\cite{Draper:1989bp,Bunton:2006va} shows
that the form factors of all isospin $I = 1$ mesons (pseudoscalar, vector, $\ldots$)
can be computed on the lattice at continuous values of the momentum transfer. 
Thus, for example, the magnetic moment and electric quadrupole moment of
the $\rho^+$ meson are accessible at zero Fourier momentum.
Apart from the pseudoscalar mesons, the volume effects cannot be calculated systematically, 
however, as there are no known EFTs for these mesons with a controlled power counting.
One would have to demonstrate that the electromagnetic properties extracted from 
the lattice are stable under changes in the volume. 
Nonetheless simulating lattice QCD with flavor twisted boundary conditions
provides a promising technique to overcome the restriction to quantized momentum 
transfer.

\begin{acknowledgments}
This work is supported in part by the U.S.\ Dept.~of Energy,
Grant No.\ DE-FG02-05ER41368-0 (F.-J.J. and B.C.T.) and by the
Schweizerischer Nationalfonds (F.-J.J.). 
\end{acknowledgments}

\appendix


\begin{thebibliography}{33}
\expandafter\ifx\csname natexlab\endcsname\relax\def\natexlab#1{#1}\fi
\expandafter\ifx\csname bibnamefont\endcsname\relax
  \def\bibnamefont#1{#1}\fi
\expandafter\ifx\csname bibfnamefont\endcsname\relax
  \def\bibfnamefont#1{#1}\fi
\expandafter\ifx\csname citenamefont\endcsname\relax
  \def\citenamefont#1{#1}\fi
\expandafter\ifx\csname url\endcsname\relax
  \def\url#1{\texttt{#1}}\fi
\expandafter\ifx\csname urlprefix\endcsname\relax\def\urlprefix{URL }\fi
\providecommand{\bibinfo}[2]{#2}
\providecommand{\eprint}[2][]{\url{#2}}

\bibitem[{\citenamefont{Bernard et~al.}(2003)}]{Bernard:2002yk}
\bibinfo{author}{\bibfnamefont{C.}~\bibnamefont{Bernard}} \bibnamefont{et~al.},
  \bibinfo{journal}{Nucl. Phys. Proc. Suppl.} \textbf{\bibinfo{volume}{119}},
  \bibinfo{pages}{170} (\bibinfo{year}{2003}), \eprint{hep-lat/0209086}.

\bibitem[{\citenamefont{Gross and Kitazawa}(1982)}]{Gross:1982at}
\bibinfo{author}{\bibfnamefont{D.~J.} \bibnamefont{Gross}} \bibnamefont{and}
  \bibinfo{author}{\bibfnamefont{Y.}~\bibnamefont{Kitazawa}},
  \bibinfo{journal}{Nucl. Phys.} \textbf{\bibinfo{volume}{B206}},
  \bibinfo{pages}{440} (\bibinfo{year}{1982}).

\bibitem[{\citenamefont{Roberge and Weiss}(1986)}]{Roberge:1986mm}
\bibinfo{author}{\bibfnamefont{A.}~\bibnamefont{Roberge}} \bibnamefont{and}
  \bibinfo{author}{\bibfnamefont{N.}~\bibnamefont{Weiss}},
  \bibinfo{journal}{Nucl. Phys.} \textbf{\bibinfo{volume}{B275}},
  \bibinfo{pages}{734} (\bibinfo{year}{1986}).

\bibitem[{\citenamefont{Wiese}(1992)}]{Wiese:1991ku}
\bibinfo{author}{\bibfnamefont{U.~J.} \bibnamefont{Wiese}},
  \bibinfo{journal}{Nucl. Phys.} \textbf{\bibinfo{volume}{B375}},
  \bibinfo{pages}{45} (\bibinfo{year}{1992}).

\bibitem[{\citenamefont{Luscher et~al.}(1996)\citenamefont{Luscher, Sint,
  Sommer, and Weisz}}]{Luscher:1996sc}
\bibinfo{author}{\bibfnamefont{M.}~\bibnamefont{Luscher}},
  \bibinfo{author}{\bibfnamefont{S.}~\bibnamefont{Sint}},
  \bibinfo{author}{\bibfnamefont{R.}~\bibnamefont{Sommer}}, \bibnamefont{and}
  \bibinfo{author}{\bibfnamefont{P.}~\bibnamefont{Weisz}},
  \bibinfo{journal}{Nucl. Phys.} \textbf{\bibinfo{volume}{B478}},
  \bibinfo{pages}{365} (\bibinfo{year}{1996}), \eprint{hep-lat/9605038}.

\bibitem[{\citenamefont{Bucarelli et~al.}(1999)\citenamefont{Bucarelli,
  Palombi, Petronzio, and Shindler}}]{Bucarelli:1998mu}
\bibinfo{author}{\bibfnamefont{A.}~\bibnamefont{Bucarelli}},
  \bibinfo{author}{\bibfnamefont{F.}~\bibnamefont{Palombi}},
  \bibinfo{author}{\bibfnamefont{R.}~\bibnamefont{Petronzio}},
  \bibnamefont{and} \bibinfo{author}{\bibfnamefont{A.}~\bibnamefont{Shindler}},
  \bibinfo{journal}{Nucl. Phys.} \textbf{\bibinfo{volume}{B552}},
  \bibinfo{pages}{379} (\bibinfo{year}{1999}), \eprint{hep-lat/9808005}.

\bibitem[{\citenamefont{Guagnelli et~al.}(2003)}]{Guagnelli:2003hw}
\bibinfo{author}{\bibfnamefont{M.}~\bibnamefont{Guagnelli}}
  \bibnamefont{et~al.} (\bibinfo{collaboration}{Zeuthen-Rome / ZeRo}),
  \bibinfo{journal}{Nucl. Phys.} \textbf{\bibinfo{volume}{B664}},
  \bibinfo{pages}{276} (\bibinfo{year}{2003}), \eprint{hep-lat/0303012}.

\bibitem[{\citenamefont{Kiskis et~al.}(2002)\citenamefont{Kiskis, Narayanan,
  and Neuberger}}]{Kiskis:2002gr}
\bibinfo{author}{\bibfnamefont{J.}~\bibnamefont{Kiskis}},
  \bibinfo{author}{\bibfnamefont{R.}~\bibnamefont{Narayanan}},
  \bibnamefont{and}
  \bibinfo{author}{\bibfnamefont{H.}~\bibnamefont{Neuberger}},
  \bibinfo{journal}{Phys. Rev.} \textbf{\bibinfo{volume}{D66}},
  \bibinfo{pages}{025019} (\bibinfo{year}{2002}), \eprint{hep-lat/0203005}.

\bibitem[{\citenamefont{Kiskis et~al.}(2003)\citenamefont{Kiskis, Narayanan,
  and Neuberger}}]{Kiskis:2003rd}
\bibinfo{author}{\bibfnamefont{J.}~\bibnamefont{Kiskis}},
  \bibinfo{author}{\bibfnamefont{R.}~\bibnamefont{Narayanan}},
  \bibnamefont{and}
  \bibinfo{author}{\bibfnamefont{H.}~\bibnamefont{Neuberger}},
  \bibinfo{journal}{Phys. Lett.} \textbf{\bibinfo{volume}{B574}},
  \bibinfo{pages}{65} (\bibinfo{year}{2003}), \eprint{hep-lat/0308033}.

\bibitem[{\citenamefont{Kim and Christ}(2003)}]{Kim:2002np}
\bibinfo{author}{\bibfnamefont{C.-H.} \bibnamefont{Kim}} \bibnamefont{and}
  \bibinfo{author}{\bibfnamefont{N.~H.} \bibnamefont{Christ}},
  \bibinfo{journal}{Nucl. Phys. Proc. Suppl.} \textbf{\bibinfo{volume}{119}},
  \bibinfo{pages}{365} (\bibinfo{year}{2003}), \eprint{hep-lat/0210003}.

\bibitem[{\citenamefont{Kim}(2004)}]{Kim:2003xt}
\bibinfo{author}{\bibfnamefont{C.-H.} \bibnamefont{Kim}},
  \bibinfo{journal}{Nucl. Phys. Proc. Suppl.} \textbf{\bibinfo{volume}{129}},
  \bibinfo{pages}{197} (\bibinfo{year}{2004}), \eprint{hep-lat/0311003}.

\bibitem[{\citenamefont{Bedaque}(2004)}]{Bedaque:2004kc}
\bibinfo{author}{\bibfnamefont{P.~F.} \bibnamefont{Bedaque}},
  \bibinfo{journal}{Phys. Lett.} \textbf{\bibinfo{volume}{B593}},
  \bibinfo{pages}{82} (\bibinfo{year}{2004}), \eprint{nucl-th/0402051}.

\bibitem[{\citenamefont{de~Divitiis et~al.}(2004)\citenamefont{de~Divitiis,
  Petronzio, and Tantalo}}]{deDivitiis:2004kq}
\bibinfo{author}{\bibfnamefont{G.~M.} \bibnamefont{de~Divitiis}},
  \bibinfo{author}{\bibfnamefont{R.}~\bibnamefont{Petronzio}},
  \bibnamefont{and} \bibinfo{author}{\bibfnamefont{N.}~\bibnamefont{Tantalo}},
  \bibinfo{journal}{Phys. Lett.} \textbf{\bibinfo{volume}{B595}},
  \bibinfo{pages}{408} (\bibinfo{year}{2004}), \eprint{hep-lat/0405002}.

\bibitem[{\citenamefont{Sachrajda and Villadoro}(2005)}]{Sachrajda:2004mi}
\bibinfo{author}{\bibfnamefont{C.~T.} \bibnamefont{Sachrajda}}
  \bibnamefont{and}
  \bibinfo{author}{\bibfnamefont{G.}~\bibnamefont{Villadoro}},
  \bibinfo{journal}{Phys. Lett.} \textbf{\bibinfo{volume}{B609}},
  \bibinfo{pages}{73} (\bibinfo{year}{2005}), \eprint{hep-lat/0411033}.

\bibitem[{\citenamefont{Bedaque and Chen}(2005)}]{Bedaque:2004ax}
\bibinfo{author}{\bibfnamefont{P.~F.} \bibnamefont{Bedaque}} \bibnamefont{and}
  \bibinfo{author}{\bibfnamefont{J.-W.} \bibnamefont{Chen}},
  \bibinfo{journal}{Phys. Lett.} \textbf{\bibinfo{volume}{B616}},
  \bibinfo{pages}{208} (\bibinfo{year}{2005}), \eprint{hep-lat/0412023}.

\bibitem[{\citenamefont{Tiburzi}(2005)}]{Tiburzi:2005hg}
\bibinfo{author}{\bibfnamefont{B.~C.} \bibnamefont{Tiburzi}},
  \bibinfo{journal}{Phys. Lett.} \textbf{\bibinfo{volume}{B617}},
  \bibinfo{pages}{40} (\bibinfo{year}{2005}), \eprint{hep-lat/0504002}.

\bibitem[{\citenamefont{Flynn et~al.}(2006)\citenamefont{Flynn, Juttner, and
  Sachrajda}}]{Flynn:2005in}
\bibinfo{author}{\bibfnamefont{J.~M.} \bibnamefont{Flynn}},
  \bibinfo{author}{\bibfnamefont{A.}~\bibnamefont{Juttner}}, \bibnamefont{and}
  \bibinfo{author}{\bibfnamefont{C.~T.} \bibnamefont{Sachrajda}}
  (\bibinfo{collaboration}{UKQCD}), \bibinfo{journal}{Phys. Lett.}
  \textbf{\bibinfo{volume}{B632}}, \bibinfo{pages}{313} (\bibinfo{year}{2006}),
  \eprint{hep-lat/0506016}.

\bibitem[{\citenamefont{Guadagnoli et~al.}(2006)\citenamefont{Guadagnoli,
  Mescia, and Simula}}]{Guadagnoli:2005be}
\bibinfo{author}{\bibfnamefont{D.}~\bibnamefont{Guadagnoli}},
  \bibinfo{author}{\bibfnamefont{F.}~\bibnamefont{Mescia}}, \bibnamefont{and}
  \bibinfo{author}{\bibfnamefont{S.}~\bibnamefont{Simula}},
  \bibinfo{journal}{Phys. Rev.} \textbf{\bibinfo{volume}{D73}},
  \bibinfo{pages}{114504} (\bibinfo{year}{2006}), \eprint{hep-lat/0512020}.

\bibitem[{\citenamefont{Aarts et~al.}(2006)\citenamefont{Aarts, Allton, Foley,
  Hands, and Kim}}]{Aarts:2006wt}
\bibinfo{author}{\bibfnamefont{G.}~\bibnamefont{Aarts}},
  \bibinfo{author}{\bibfnamefont{C.}~\bibnamefont{Allton}},
  \bibinfo{author}{\bibfnamefont{J.}~\bibnamefont{Foley}},
  \bibinfo{author}{\bibfnamefont{S.}~\bibnamefont{Hands}}, \bibnamefont{and}
  \bibinfo{author}{\bibfnamefont{S.}~\bibnamefont{Kim}} (\bibinfo{year}{2006}),
  \eprint{hep-lat/0607012}.

\bibitem[{\citenamefont{Tiburzi}(2006)}]{Tiburzi:2006px}
\bibinfo{author}{\bibfnamefont{B.~C.} \bibnamefont{Tiburzi}},
  \bibinfo{journal}{Phys.~Lett.} \textbf{\bibinfo{volume}{B641}},
  \bibinfo{pages}{342} (\bibinfo{year}{2006}), \eprint{hep-lat/0607019}.

\bibitem[{\citenamefont{Martinelli and Sachrajda}(1988)}]{Martinelli:1988bh}
\bibinfo{author}{\bibfnamefont{G.}~\bibnamefont{Martinelli}} \bibnamefont{and}
  \bibinfo{author}{\bibfnamefont{C.~T.} \bibnamefont{Sachrajda}},
  \bibinfo{journal}{Nucl. Phys.} \textbf{\bibinfo{volume}{B306}},
  \bibinfo{pages}{865} (\bibinfo{year}{1988}).

\bibitem[{\citenamefont{Draper et~al.}(1989)\citenamefont{Draper, Woloshyn,
  Wilcox, and Liu}}]{Draper:1989bp}
\bibinfo{author}{\bibfnamefont{T.}~\bibnamefont{Draper}},
  \bibinfo{author}{\bibfnamefont{R.~M.} \bibnamefont{Woloshyn}},
  \bibinfo{author}{\bibfnamefont{W.}~\bibnamefont{Wilcox}}, \bibnamefont{and}
  \bibinfo{author}{\bibfnamefont{K.-F.} \bibnamefont{Liu}},
  \bibinfo{journal}{Nucl. Phys.} \textbf{\bibinfo{volume}{B318}},
  \bibinfo{pages}{319} (\bibinfo{year}{1989}).

\bibitem[{\citenamefont{Bunton et~al.}(2006)\citenamefont{Bunton, Jiang, and
  Tiburzi}}]{Bunton:2006va}
\bibinfo{author}{\bibfnamefont{T.~B.} \bibnamefont{Bunton}},
  \bibinfo{author}{\bibfnamefont{F.~J.} \bibnamefont{Jiang}}, \bibnamefont{and}
  \bibinfo{author}{\bibfnamefont{B.~C.} \bibnamefont{Tiburzi}},
  \bibinfo{journal}{Phys. Rev.} \textbf{\bibinfo{volume}{D74}},
  \bibinfo{pages}{034514} (\bibinfo{year}{2006}), \eprint{hep-lat/0607001}.

\bibitem[{\citenamefont{Bonnet et~al.}(2005)\citenamefont{Bonnet, Edwards,
  Fleming, Lewis, and Richards}}]{Bonnet:2004fr}
\bibinfo{author}{\bibfnamefont{F.~D.~R.} \bibnamefont{Bonnet}},
  \bibinfo{author}{\bibfnamefont{R.~G.} \bibnamefont{Edwards}},
  \bibinfo{author}{\bibfnamefont{G.~T.} \bibnamefont{Fleming}},
  \bibinfo{author}{\bibfnamefont{R.}~\bibnamefont{Lewis}}, \bibnamefont{and}
  \bibinfo{author}{\bibfnamefont{D.~G.} \bibnamefont{Richards}}
  (\bibinfo{collaboration}{LHP}), \bibinfo{journal}{Phys.
  Rev.} \textbf{\bibinfo{volume}{D72}}, \bibinfo{pages}{054506}
  (\bibinfo{year}{2005}), \eprint{hep-lat/0411028}.

\bibitem[{\citenamefont{Hashimoto et~al.}(2006)}]{Hashimoto:2005am}
\bibinfo{author}{\bibfnamefont{S.}~\bibnamefont{Hashimoto}}
  \bibnamefont{et~al.} (\bibinfo{collaboration}{JLQCD}), \bibinfo{journal}{PoS}
  \textbf{\bibinfo{volume}{LAT2005}}, \bibinfo{pages}{336}
  (\bibinfo{year}{2006}), \eprint{hep-lat/0510085}.

\bibitem[{\citenamefont{"Brommel et~al.}(2006)}]{Brommel:2006ww}
\bibinfo{author}{\bibfnamefont{D.}~\bibnamefont{Brommel}}
  \bibnamefont{et~al.} (\bibinfo{collaboration}{QCDSF/UKQCD}),
\eprint{hep-lat/0608021}.


\bibitem[{\citenamefont{Sharpe and Shoresh}(2000)}]{Sharpe:2000bc}
\bibinfo{author}{\bibfnamefont{S.~R.} \bibnamefont{Sharpe}} \bibnamefont{and}
  \bibinfo{author}{\bibfnamefont{N.}~\bibnamefont{Shoresh}},
  \bibinfo{journal}{Phys. Rev.} \textbf{\bibinfo{volume}{D62}},
  \bibinfo{pages}{094503} (\bibinfo{year}{2000}),
  \eprint[http://arXiv.org/abs]{hep-lat/0006017}.

\bibitem[{\citenamefont{Gasser and Leutwyler}(1988)}]{Gasser:1987zq}
\bibinfo{author}{\bibfnamefont{J.}~\bibnamefont{Gasser}} \bibnamefont{and}
  \bibinfo{author}{\bibfnamefont{H.}~\bibnamefont{Leutwyler}},
  \bibinfo{journal}{Nucl. Phys.} \textbf{\bibinfo{volume}{B307}},
  \bibinfo{pages}{763} (\bibinfo{year}{1988}).

\bibitem[{\citenamefont{Bernard and Golterman}(1994)}]{Bernard:1994sv}
\bibinfo{author}{\bibfnamefont{C.~W.} \bibnamefont{Bernard}} \bibnamefont{and}
  \bibinfo{author}{\bibfnamefont{M.~F.~L.} \bibnamefont{Golterman}},
  \bibinfo{journal}{Phys. Rev.} \textbf{\bibinfo{volume}{D49}},
  \bibinfo{pages}{486} (\bibinfo{year}{1994}),
  \eprint[http://arXiv.org/abs]{hep-lat/9306005}.

\bibitem[{\citenamefont{Sharpe}(1997)}]{Sharpe:1997by}
\bibinfo{author}{\bibfnamefont{S.~R.} \bibnamefont{Sharpe}},
  \bibinfo{journal}{Phys. Rev.} \textbf{\bibinfo{volume}{D56}},
  \bibinfo{pages}{7052} (\bibinfo{year}{1997}),
  \eprint[http://arXiv.org/abs]{hep-lat/9707018}.

\bibitem[{\citenamefont{Golterman and Leung}(1998)}]{Golterman:1998st}
\bibinfo{author}{\bibfnamefont{M.~F.~L.} \bibnamefont{Golterman}}
  \bibnamefont{and} \bibinfo{author}{\bibfnamefont{K.-C.} \bibnamefont{Leung}},
  \bibinfo{journal}{Phys. Rev.} \textbf{\bibinfo{volume}{D57}},
  \bibinfo{pages}{5703} (\bibinfo{year}{1998}),
  \eprint[http://arXiv.org/abs]{hep-lat/9711033}.


\bibitem[{\citenamefont{Sharpe and Shoresh}(2001)}]{Sharpe:2001fh}
\bibinfo{author}{\bibfnamefont{S.~R.} \bibnamefont{Sharpe}} \bibnamefont{and}
  \bibinfo{author}{\bibfnamefont{N.}~\bibnamefont{Shoresh}},
  \bibinfo{journal}{Phys. Rev.} \textbf{\bibinfo{volume}{D64}},
  \bibinfo{pages}{114510} (\bibinfo{year}{2001}),
  \eprint[http://arXiv.org/abs]{hep-lat/0108003}.


\bibitem[{\citenamefont{Zinn-Justin}(2002)}]{Zinn-Justin:2002ru}
\bibinfo{author}{\bibfnamefont{J.} \bibnamefont{Zinn-Justin}}, 
\bibinfo{title}{{\it Quantum Field Theory and Critical Phenomena}},
  \bibinfo{journal}{Int. Ser. Monogr. Phys.} \textbf{\bibinfo{volume}{113}},
  \bibinfo{pages}{1-1054} (\bibinfo{year}{2002}).


\bibitem[{\citenamefont{Gasser and Leutwyler}(1984)}]{Gasser:1983yg}
\bibinfo{author}{\bibfnamefont{J.}~\bibnamefont{Gasser}} \bibnamefont{and}
  \bibinfo{author}{\bibfnamefont{H.} \bibnamefont{Leutwyler}},
  \bibinfo{journal}{Ann.~Phys.} \textbf{\bibinfo{volume}{158}},
  \bibinfo{pages}{142} (\bibinfo{year}{1984}).


\bibitem[{\citenamefont{Gasser and Leutwyler}(1985)}]{Gasser:1984ux}
\bibinfo{author}{\bibfnamefont{J.}~\bibnamefont{Gasser}} \bibnamefont{and}
  \bibinfo{author}{\bibfnamefont{H.} \bibnamefont{Leutwyler}},
  \bibinfo{journal}{Nucl.~Phys.} \textbf{\bibinfo{volume}{B250}},
  \bibinfo{pages}{517} (\bibinfo{year}{1985}).

\bibitem[{\citenamefont{Donoghue}(1992)}]{Donoghue:1992dd}
\bibinfo{author}{\bibfnamefont{J.~F.} \bibnamefont{Donoghue}, \bibnamefont{E.} \bibnamefont{Golowich}, \bibnamefont{and}
\bibnamefont{B.~R.} \bibnamefont{Holstein}}, 
\bibinfo{title}{{\it Dynamics of the Standard Model}},
  \bibinfo{journal}{Camb. Monogr. Part. Phys. Nucl. Phys. Cosmol.} \textbf{\bibinfo{volume}{2}},
  \bibinfo{pages}{1-540} (\bibinfo{year}{1992}).


\bibitem[{\citenamefont{Arndt and Tiburzi}(2003)}]{Arndt:2003ww}
\bibinfo{author}{\bibfnamefont{D.}~\bibnamefont{Arndt}} \bibnamefont{and}
  \bibinfo{author}{\bibfnamefont{B.~C.} \bibnamefont{Tiburzi}},
  \bibinfo{journal}{Phys. Rev.} \textbf{\bibinfo{volume}{D68}},
  \bibinfo{pages}{094501} (\bibinfo{year}{2003}), \eprint{hep-lat/0307003}.

\bibitem[{\citenamefont{Detmold and Lin}(2005)}]{Detmold:2005pt}
\bibinfo{author}{\bibfnamefont{W.}~\bibnamefont{Detmold}} \bibnamefont{and}
  \bibinfo{author}{\bibfnamefont{C.~J.~D.} \bibnamefont{Lin}},
  \bibinfo{journal}{Phys. Rev.} \textbf{\bibinfo{volume}{D71}},
  \bibinfo{pages}{054510} (\bibinfo{year}{2005}), \eprint{hep-lat/0501007}.

\end{thebibliography}

\end{document}